# Reduction of circuit depth by mapping qubit-based quantum gates to a qudit basis

*Pamela Rambow[1] and Mingzhen Tian[2]*
*Department of Physics and Astronomy, Quantum Science and Engineering Center*
*George Mason University, Fairfax, Virginia 22030*

We present a scalable set of universal and multiply controlled gates in a qudit basis through a bijective mapping from N qubits to qudits with $D = 2^N$ levels via rotations in $U(2)$. For each of the universal gates (*H*, *CNOT*, and *T*), as well as the *NOT* gate and multiply-controlled-*Z* gates, we describe a systematic approach to identifying the set of $U(2)$ rotations required to implement each gate for any qudit of size D and with minimal use of an ancilla level. The qudit gates are analyzed in terms of the total rotation count and gate depth as the system scales with $N$. We apply the qudit-basis gates to Grover's Algorithm and compare the circuit depth vs. system size to a qubit-based circuit. The results show that there is a dramatic reduction in circuit depth as the size of the system increases for the qudit circuit compared to qubit circuit. In particular, multiply controlled gates are the driving factor in the reduction of circuit complexity for qudit-based circuits since the gate depth remains constant as the system scales with D.

## I. INRODUCTION

The development of quantum algorithms and implementation has been focused on qubit-based systems [1-9]. A quantum algorithm is implemented on quantum processors through a quantum circuit constructed with a set of universal gates, such as, the commonly used Hadamard (*H*), pi/8-phase (*T*), and controlled-not (*CNOT*) gates. The success of the qubit-based circuit model of quantum computation is rooted in its universality realized through "local" gates operating on single or two qubits [6,10]. On the hardware level, these gates are composed with finite rotation operators (or conditional rotation operators in *CNOT* gate) in the 2-dimensional Hilbert space of a qubit driven by external Hamiltonians.

Optimization of quantum circuits have attracted extensive efforts due to its significance in studying theoretical complexity bounds for a given algorithm. It is also critical for implementing an algorithm successfully on hardware available currently or in the near future. For practical reasons, circuit complexity analysis usually includes two parts: the total gate count that can be either the basic universal gates or rotation operators in a circuit, and the circuit depth that counts the number of gates (or rotations) on the longest gate sequence in the circuit. In a noisy intermediate-scale quantum computer, limited coherence time sets an upper bound for the circuit depth and errors in quantum gates accumulates with the total number of gates [11]. In addition, two-qubit gates, such as *CNOT* and controlled-Z (*CZ*) gates, are currently the dominant sources of error due to their higher error rates compared with single qubit gates [12].

Multiple-qubit gates, such as Toffoli and doubly-controlled-Z gates and their multiply-controlled versions, need complex circuits to implement in a qubit-based system. A Toffoli gate with $N-1$ controls and one target qubit can be implemented with an *N*-qubit circuit of quadratic depth and gate count $\mathcal{O}(N^2)$[13]. The complexity can be reduced to $\mathcal{O}(N)$ with the help of at least one ancillary qubit [13-15]. It has also been proven that such a qubit-based circuit needs at least 2*N* *CNOT* gates [16]. The theoretical lower bound of the circuit depth is still unknown. However, the existing schemes with the least depth of $\mathcal{O}(\log N)$ requires extra resources including $\mathcal{O}(N)$ clean ancillary qubits and disentangling operations conditional on measurement results [15].

A quantum computer using one or more qudits rather than an all-qubit system may provide a viable solution in reducing circuit complexity, especially when involving multiple-qubit gates as mentioned above [17-32].

In this paper, we will focus on methods for mapping qubit-specified gates for an arbitrary *N*-qubit system to one qudit of dimension $D = 2^N + 1$, where the extra level is needed in some cases. The universal gate set, as well as some important multiply-controlled gates and combinations of multiple single qubit gates applied in parallel will be considered. We assume the computational resources available are rotations in 2-dimensional subspaces of the qudit with fixed angles in integer number of $\pi/2$. Circuit complexity is analyzed in term of the rotation operators. Our results show significant reduction in both total rotation count and circuit depth in the $(N-1)$-controlled-Z gate that can be constructed with a single $2\pi$ rotation independent of $N$. Circuit depth for a single qudit gate remains constant while the total rotation count increases in $\mathcal{O}(D)$. Furthermore, the complexity of the qudit gate remains the same when an equivalent single qubit gate, such as *T* or *NOT* gate is applied to multiple qubits in parallel. The most complex gate is the *H* gate. When applied to multiple qubits, the rotation count grows in $\mathcal{O}(DN)$ and depth, in $\mathcal{O}(N)$ for the equivalent qudit *H* gate. We will apply these features to show the reduction of circuit complexity, both in overall rotation counts and the circuit depth in Grover's algorithm implemented in a qudit.

---

[1] prambow@gmu.edu
[2] mtian@gmu.edu

The rest of the paper is arranged in three main sections. In Sec. IIA we first lay the framework to express qudit gates in a qubit basis along with the generalized construction of qudit gates via unitary rotations in the $U(2)$ subspaces. Next in Sec IIB, we describe the specific sequence of rotations to construct the universal gate set, some commonly used gate combinations, as well multiply-controlled-Z gates. In Sec. III we implement Grover's algorithm using qudit gates and analyze the complexity of the circuits in a qudit vs. qubit systems. In Sec. VI we summarize our results and discuss prospective work.

## II. QUANTUM GATES

### A. Qubit-specified gates in qudit basis

The state vector for an arbitrary qudit of dimension $D$ is a linear superposition of the orthonormal basis of the qudit, $\{|d\rangle_D\}_{d \in \{0...D-1\}}$,

$$|\phi\rangle = \sum_{d=0}^{D-1} \alpha_d |d\rangle_D,$$

where $\alpha_d$ denotes a set of complex probability amplitudes. The subscript $D$ outside the basis ket indicates the total size of the system. For simplicity, we will restrict to $D = 2^N$ which is equivalent to a system of N qubits. The basis vector mapping between a $D$-dimensional qudit and N qubits is the tensor product of the individual qubit basis,

$$|d\rangle_D = |q_1 q_2 \ldots q_i \ldots q_N\rangle, \quad (1)$$

where the $i \in \{1,2,\ldots,N\}$ and $q_i \in \{0,1\}$. The transformation between the two sets of bases is connected through the integer $d$ and its matching Boolean bit string $q_1 q_2 \ldots q_i \ldots q_N$.

The qudit construction of equivalent one- and two-qubit gates are generally "single action" gates. The action of the qudit operator in effect equivalently transforms a single target qubit $t$,

$$U^t |d\rangle_D = U^t |q_1 \ldots q_t \ldots q_N\rangle = |q_1 \ldots q'_t \ldots q_N\rangle.$$

For a two-qubit controlled gate the nomenclature is $U^{c,t}$, where $c$ and $t$ denote the control and target qubits, respectively.

The qubit-based gates are implemented by $U(2)$ rotations in a 2-dimensional space spanned by the qubit basis. The generalized expression for a $U(2)$ rotation in the sub-space of a qudit is given by,

$$R_{\hat{n}}^{(j,k)}(\theta) = exp\left(\frac{-i\theta(\boldsymbol{\sigma} \cdot \hat{\boldsymbol{n}})}{2}\right)^{(j,k)} \oplus \mathbb{I}_{D-1}^{\overline{(j,k)}}, \quad (2)$$

Where $\boldsymbol{\sigma} \cdot \hat{\boldsymbol{n}}$ is the Pauli vector, $\hat{n}$ and $\theta$ are the axis and angle of rotation, respectively, and $j < k \in \{0,1,2,\ldots,D\}$ are the qudit levels under rotation. The first $D$ levels are the qudit basis vectors in Equation (1) and the $Dth$ is an ancillary level. The ancillary level is included to ensure the rotation in $U(2)$ does not impart a local phase to the qudit basis state. It should be noted that not all gates require the ancilla level.

Any qudit unitary can be constructed through a sequence of $U(2)$ rotations of selected axes, rotation angles, and pairings of qudit levels. The complexity analysis of a gate construction includes the number of the rotations and the depth of the circuit in terms of rotation count on the longest sequence. Our goal is to reduce the rotation count and the circuit depth. Commutation properties among the rotation operators helps to achieve reduction in both. Two useful commuting rotations are

1. Rotations operating on entirely different qudit levels commute,

2. Two rotations which share a common level commute if they share a set of eigenvectors in the Hilbert space spanned by all levels.

In the following sections we present constructions of the three universal gates (*H*, *CNOT*, and *T*), as well as *NOT* and multiply-controlled-Z gates, for a qudit system of dimension $D = 2^N$ using the qudit rotation operator (Equation 2). We will show it is sufficient to set the axes of rotations to x, y, and z, and limit rotation angles to $\pm n\pi/2$ for integer $n$.

## B. Universal Gate Set

### Hadamard Gate

The single-action qudit Hadamard ($H$) gate, $H^t$, applied on a qudit basis creates a superposition between paired qudit levels is equivalent to a qubit H gate applied to the $t$-th qubit,

$$H^t|d\rangle_D = H^t|q_1 q_2 \ldots q_t \ldots q_N\rangle$$
$$= |q_1 q_2 \ldots\rangle \otimes H|q_t\rangle \otimes |\ldots q_N\rangle.$$

The operator $H^t$ transforms every pair as

$$H^t|l\rangle_D = \frac{1}{\sqrt{2}}(|l\rangle_D + |u\rangle_D)$$
$$H^t|u\rangle_D = \frac{1}{\sqrt{2}}(|l\rangle_D - |u\rangle_D),$$

where the pairs are identified by lower and upper indices, $l$ and $u$, with $q_t = 0$ for the lower state and $q_t = 1$ for the upper state of the target qubit,

$$|l\rangle_D = |q_1 q_2 \ldots q_t = 0 \ldots q_N\rangle$$
$$|u\rangle_D = |q_1 q_2 \ldots q_t = 1 \ldots q_N\rangle.$$

For $D = 2^N$ levels there are $D/2$ pairs. The integers that label the paired levels are related by $u - l = 2^{N-t}$.

Any operator $H^t$ can be implemented by a set of $D/2$ y-rotations, $R_y(-\pi/2)$, followed by a set of $D/4$ x-rotations, $R_x(2\pi)$. Each y-rotation applied to a paired $|l\rangle_D$ and $|u\rangle_D$ creates the desired superposition but leaves a $e^{i\pi}$ phase on $|u\rangle_D$. The same phase affecting all $D/2$ upper levels can be corrected by $R_x(2\pi)$ rotations applied to $D/4$ pairs of upper levels with any pairing combinations. Therefore, the total rotation count for any single action gate $H^t$ is $3D/4$. The total circuit depth is 2 since all y-rotations commute, as well as all x-rotations, and can be applied in parallel. The two rotation sets are commutable as well.

An example of the level pairings and rotation operators for a single-action qudit $H$ gate, equivalent to a qubit $H$ gate acting on the $t = 1$ qubit, for a $D = 8$ qudit system is shown in Table 1. The rotations in curly brackets are applied in parallel.

It is quite common in an algorithm that a set of qubit $H$ gates are applied in parallel to a few or all the qubits in the circuit. Any combination of parallel qubit $H$ gates can be expressed as a multi-action qudit $H$ gate, performing the equivalent transformation. Figure 1 shows the rotation sequence for a D = 8 qudit multi-action $H$ gate equivalent to implementing qubit $H$ gates to all three qubits in parallel. The first set of y-rotations is applied to the paired levels corresponding to $t = 1$, the second and third for $t = 2$ and 3, respectively. The phase correction is done at the end by the set of x-rotations applied to $D/2$ upper states that have been addressed an odd number of times in the y-rotations.

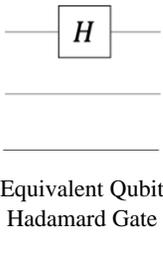

| | |
|---|---|
| | $H^{t=1}|d\rangle_8 = \frac{1}{\sqrt{2}}(|0\rangle \pm |1\rangle) \otimes |q_2 q_3\rangle$ |
| Equivalent Qubit Hadamard Gate | $H^{t=1}|0\rangle_8 = \frac{1}{\sqrt{2}}(|0\rangle_8 + |4\rangle_8)$ |
| | $H^{t=1}|1\rangle_8 = \frac{1}{\sqrt{2}}(|1\rangle_8 + |5\rangle_8)$ |
| | $H^{t=1}|2\rangle_8 = \frac{1}{\sqrt{2}}(|2\rangle_8 + |6\rangle_8)$ |
| | $H^{t=1}|3\rangle_8 = \frac{1}{\sqrt{2}}(|3\rangle_8 + |7\rangle_8)$ |
| | $H^{t=1}|4\rangle_8 = \frac{1}{\sqrt{2}}(|0\rangle_8 - |4\rangle_8)$ |
| | $H^{t=1}|5\rangle_8 = \frac{1}{\sqrt{2}}(|1\rangle_8 - |5\rangle_8)$ |
| | $H^{t=1}|6\rangle_8 = \frac{1}{\sqrt{2}}(|2\rangle_8 - |6\rangle_8)$ |
| | $H^{t=1}|7\rangle_8 = \frac{1}{\sqrt{2}}(|3\rangle_8 - |7\rangle_8)$ |
| $H^{t=1} =$ | $\{R_x^{(4,6)}(2\pi) R_x^{(5,7)}(2\pi)\}$ $* \{R_y^{(0,4)}\left(\frac{-\pi}{2}\right) R_y^{(1,5)}\left(-\frac{\pi}{2}\right) R_y^{(2,6)}\left(\frac{-\pi}{2}\right) R_y^{(3,7)}\left(\frac{-\pi}{2}\right)\}$ |

Table I. Qudit Hadamard gate example for a $D = 8$ system with the $H$ gate action on the $t = 1$ equivalent qubit. The rotation operators reads from right to left.

For an arbitrary number of $M$ qubit $H$ gates applied in parallel, where $1 \leq M \leq N$, the equivalent multi-action qudit $H$ gate is constructed by $M$ sets of y-rotations, each set applied to the paired levels corresponding to the target qubits, and $D/4$ sets of phase correction x-rotations.

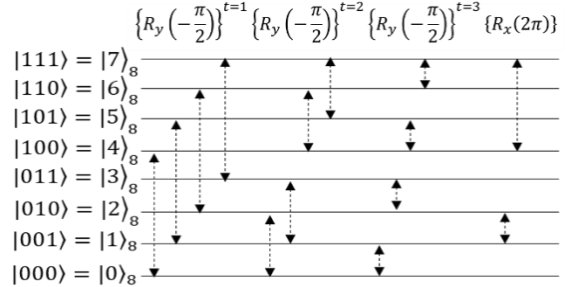

Figure 1. An $H$ gate with action on all equivalent qubits for a $D = 8$ qudit system requires a total of 14 rotations. The dashed double-sided arrows indicate the level pairings addressed by each rotation. Each set of rotations in curly brackets commute and can be performed in parallel. The operator sequence reads from left to right.

For the multi-action qudit $H$ gate, the total rotation count is $D(2M + 1)/4$. All $R_y(-\pi/2)$ rotations for a given $t$ commute, as do all the $R_x(2\pi)$ rotations. Therefore, the circuit depth scales as $M + 1$, i.e. it scales by the number of equivalent parallel qubit $H$ gates, with the upper limit for the circuit depth when $M = N$.

*CNOT Gate*

The qudit *CNOT* gate performs the same action as an equivalent qubit *CNOT* gate where the target state is flipped predicated on the state of the control qubit,

$$C_x^{c,t}|d\rangle_D = C_x^{c,t}|q_1 \dots q_c \dots q_t \dots q_N\rangle$$
$$= |q_1 \dots q \dots (q_c \oplus q_t) \dots q_N\rangle.$$

This gate can be implemented with a set of rotations in $U(2)$ subspaces,

$$C_x^{c,t} = \prod \{R_x^{b,c}(2\pi)\}\{R_y^{a,b}(\pi)\},$$

where the indices $a, b,$ and $c$ are chosen according to a given pair of control and target qubits. The rotation operator reads from right to left.

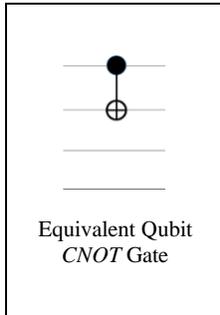

| | |
|---|---|
| | $|d\rangle_{16} = |q_c q_t q_3 q_4\rangle$ <br> $= |q_c q_t\rangle \otimes |q_3 q_4\rangle$ |
| | $|15\rangle_{16} = |1111\rangle = |11\rangle \otimes |11\rangle$ <br> $|11\rangle_{16} = |1011\rangle = |10\rangle \otimes |11\rangle$ |
| | $|14\rangle_{16} = |1110\rangle = |11\rangle \otimes |10\rangle$ <br> $|10\rangle_{16} = |1010\rangle = |10\rangle \otimes |10\rangle$ |
| Equivalent Qubit CNOT Gate | $|13\rangle_{16} = |1101\rangle = |11\rangle \otimes |01\rangle$ <br> $|9\rangle_{16} = |1001\rangle = |10\rangle \otimes |01\rangle$ |
| | $|12\rangle_{16} = |1100\rangle = |11\rangle \otimes |00\rangle$ <br> $|8\rangle_{16} = |1000\rangle = |10\rangle \otimes |00\rangle$ |
| $C_x^{1,2} = \{R_x^{(15,14)}(2\pi)R_x^{(13,12)}(2\pi)\} *$ <br> $\{R_y^{(11,15)}(\pi)R_y^{(10,14)}(\pi)R_y^{(9,13)}(\pi)R_y^{(8,12)}(\pi)\}$ | |

Table II. *CNOT* gate example of the paired levels for the $\pi$ rotation about the y-axis for a qudit of dimension $D = 16$. For each level pairing the control qubit is 1, the target qubit is either 0 or 1, and the remaining qubits are in an identical separable state. The rotation operators reads from right to left.

Table II. demonstrates an example of the $D = 16$ qudit CNOT gate, $C_x^{1,2}$, in a qubit circuit and the qubit-to-qudit mapping used to identify the rotation levels. A $R_y(\pi)$ rotation pairs the states where the target qubit is in $|1\rangle$ and $|0\rangle$, the control qubit is $|1\rangle$, and the remainder of the qubits are in identical separable states. There are $D/4$ such pairs, as listed in the table. These y-rotations flip the paired levels and impart a $e^{i\pi}$ phase to the control and target qubit both in the upper $|1\rangle$ state. The local phase is removed by a set of $R_x(2\pi)$ rotations between two of the control and target qubits in the upper state. There are $D/8$ $R_x(2\pi)$ rotations for $D \geq 8$. For $D = 4$, where the local phase is present in only one level an ancillary level is needed to compensate the local phase.

The total rotation count is $3D/8$. All $R_y(\pi)$ rotations can be run in parallel, as well as all $R_x(2\pi)$. Therefore, the circuit depth is 2 and is independent of the system size.

A multiple control qudit *CNOT* gate rotation sequence can be construction based on the same principles, i.e. the level pairing for $R_y(\pi)$ is between the states where the target qubits are in $|1\rangle$ and $|0\rangle$, all control qubits are in $|1\rangle$, and the remaining no-action qubits in an identical qubit basis state. The number of pairs equivalent to the number of orthogonal basis of the non-action qubits. For $C$ control qubits and 1 target in a $D = 2^N$ qudit, there are $2^{N-C-1}$ pairs, which is also the number of $R_y(\pi)$ rotations. In a system with at least one non-action qubit, the local phase in each pair can be compensated by a $R_x(2\pi)$ between two pairs as discussed above. The total operator count is $3(2^{N-C-1})/2$ for $N - C \geq 2$. In the cases where all qubits are involved, i.e. $C = N - 1$, the multiple *CNOT* can be implemented with one $R_y(\pi)$ and followed by a $R_x(2\pi)$ to an ancillary level.

Implementing a multiply controlled CNOT gate logic in a qubit system requires gate decomposition that grows with the number of controls from at best $\mathcal{O}(N)$ [6, 13, 16] to $\mathcal{O}(log N)$ [15, 33]. In a qudit system using the method described for a multiply controlled CNOT gate, the depth growth is reduced to $\mathcal{O}(1)$, with a constant depth of 2 and is independent of the number of controls.

*T Gate*

The $T$ gate imparts a phase of $exp(i\pi/4)$ to the qudit levels corresponding to the target qubit of $|1\rangle$,

$$T^t|d\rangle_D = T^t|q_1 \dots q_t \dots q_N\rangle$$
$$= exp(i\pi q_t/4)|q_1 \dots q_t \dots q_N\rangle.$$

There are $D/2$ levels, each requires one rotation about the z-axis by $-\pi/2$ between the corresponding level and the ancillary level. The generalized rotations sequence is,

$$T = \prod \{R_z^{(a,D)}\left(\frac{-\pi}{2}\right)\},$$

where index $a$ includes $D/2$ levels of $q_t = 1$. Table III shows an example of a $T$ gate applied to the $t = 2$ equivalent qubit.

The depth of the $T$ gate is $\mathcal{O}(1)$ with a constant depth of 1 since all $R_z(-\pi/2)$ rotations commute and can be executed in parallel. This gate can also be constructed with the same number of $R_z(-\pi/4)$ rotations without ancillary level.

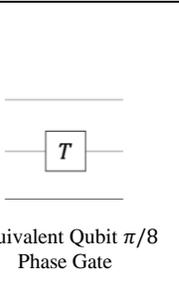

Table III. $T$ gate example in a qudit of size $D = 8$ where the qubit equivalent $T$ gate is applied to the second qubit in the circuit. The rotation operators reads from right to left.

### C. Additional Gates

In a similar manner presented in the previous section, any single-qubit and two-qubit gate can be implemented in a qudit system with rotations in $U(2)$ subspaces. In most of the cases, a set of rotations are needed to compensate unintended local phases, sometimes via an ancillary level. Implementation of multiple gates on different qubits and multiply-controlled gates may be significantly simplified in a qudit basis. An example in each case will be given in this section.

*NOT Gate*

A single qubit NOT gate is a Pauli-X operator that flips the target qubit state between $|0\rangle$ and $|1\rangle$,

$$X^t|d\rangle_D = |q_1 q_2 \ldots\rangle \otimes X|q_t\rangle \otimes |\ldots q_N\rangle.$$

In a qudit, the state pairing is similar to the *CNOT* gate. A pair consists of two levels where the target qubit state in $|0\rangle$ and $|1\rangle$ and the remaining qubits are in an identical basis state. There are $D/2$ pairs in the system, which requires $D/2$ $R_y(\pi)$ and $D/4$ $R_x(2\pi)$ rotations. All $R_y(\pi)$ rotations can be run in parallel, as well as all $R_x(2\pi)$. The circuit depth for the qudit $X$ gate has a circuit depth growth of $\mathcal{O}(1)$ with a constant depth of 2, same as in the *CNOT* gate.

Table IV demonstrates the rotation operations for a $X$ gate in $D = 8$ qudit system where the equivalent qubit circuit consist of two $X$ gates in parallel in a three qubit system. The same circuit logic can be applied using a single qudit $X$ gate requiring 6 rotations to construct the gate. Identifying the rotation level pairings for any combination of $X$ gates is a simple matter of determining the bijective mapping of the initial qudit state to the state after transformation under the gate logic. A mapping example is shown in Table IV where $X$ gates are applied to the 1st and 2nd qubits. The total numbers of $R_y(\pi)$ and $R_x(2\pi)$ rotations for a single qudit $X$ gate, which implements a set of parallel single-qubit $X$ gates, is $D/2$ and $D/4$, respectively and is independent of the number of single-qubit $X$ gates.

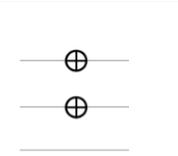

Table IV. $X$ gate example of the paired levels for a qudit system of size $D = 8$. The level pairings for the y-axis rotations are between the qudit input/output states, while the x-axis rotations correct for the phase for the upper-level states. The rotation operators reads from right to left.

*Multiply-Controlled-Z Gate*

The qudit multiply-controlled-$Z$ gate imparts a phase of $exp(i\pi) = -1$ to the states where, in the qubit basis, the target qubit and the control qubits are $|1\rangle$. In a $D = 2^N$ dimensional qudit with $N - 1$ control qubits and one target, a controlled-$Z$ gate imparts a $e^{i\pi}$ phase on a single level,

$$|1\rangle^{\otimes D} = |D - 1\rangle \to -|1\rangle^{\otimes D} = -|D - 1\rangle.$$

This is the same transformation for any combination of the control and target assignment. A single rotation of $2\pi$ about the z-axis between this level and the ancillary is all that is required to construct a qudit multiply-controlled-$Z$ gate,

$$C_Z^{D-1} = R_z^{(D-1,D)}(2\pi).$$

In fact, the rotation axis can be in any direction in the $U(2)$ subspace spanned by the qudit basis vectors $|D - 1\rangle_D$ and $|D\rangle_D$ since $R_z(2\pi) = R_y(2\pi) = R_x(2\pi)$. A multiply-controlled-$Z$ gate is the core operation in many quantum algorithms, such as Grover's algorithm. Implementing it in a qudit basis can simplify a quantum circuit significantly over the conventional qubit basis operation.

Where the total number of action qubits is less than the size of the system, there are more than one qudit levels affected by the controlled-$Z$ gates, which are the levels corresponding to all action qubits in $|1\rangle$. There are $2^{N-C-T}$ such levels for C control qubits and T target qubits. The phase can be implemented by $2\pi$ rotations between any pair of these levels, that is a total $2^{N-C-T-1}$ rotations. Table V shows an example of a $Z$ gate controlled by two qubits in a qudit system of size $D = 16$.

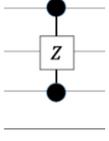

| Equivalent Qubit Multiply-Controlled-$Z$ Gate | $\|q_c q_t q_c q_4\rangle$ |
|---|---|
| | $\|15\rangle_{16} = \|1111\rangle$ <br> $\|14\rangle_{16} = \|1110\rangle$ |
| $C_z^{c,t} = R_z^{(15,14)}(2\pi)$ | |

Table V. The rotations to construct a qudit multiply-controlled-$Z$ gate, equivalent to the qubit gate diagram shown, requires two rotations by $2\pi$ about the z-axis. The rotation levels are between the ancilla and the states where the controlled and target inputs are 1. For a $D = 16$ qudit system there are two states which satisfy the conditions.

The depth of the qudit multiply-controlled $Z$ gate is 1, regardless of the size of the system, or the number of rotations to construct the qudit gate. This is because all rotations commute.

## III. GROVER'S ALGORITHM

In this section we analyze the circuit complexity of the Grover's Algorithm with a qudit and compare the circuit depth, as a function of equivalent system size for a qubit-based circuit [13, 34-36]. Whether using qubits or qudits, Grover's algorithm is implemented through the same sequence of circuit modules presented by the blocks in Figure 2 [37, 38].

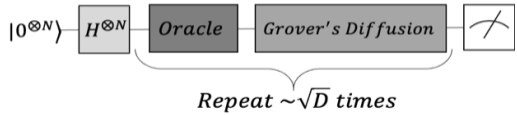

Figure 2. Grover's algorithm functional diagram for qubits and qudits. $H^{\otimes N}(2|0\rangle\langle0| - I_D)H^{\otimes N}$ is the Grover's Diffusion operator. [34]

The first step is to initiate the qudit (or qubits) from the ground state into a superposition of all basis vectors using a Hadamard gate. For a qubit system, the Hadamard gate, which has a depth of 2, is applied in parallel to each qubit in the circuit. For a qudit system, a single multi-action Hadamard gate is applied to the qudit. Recall the qudit multi-action Hadamard gate depth grows linearly as $N + 1$.

In the superposition, each basis vector represents an entry in the search list of size $D$ with one desired entry in the list to be "marked" by the oracle. The oracle module flips the sign of the probability amplitude for the term representing the marked entry while leaving the rest of system unchanged. In the construction of this module, the simplest case for the oracle is the $(N - 1)$-controlled-$Z$ gate which specifies the state $|D - 1\rangle = |1\rangle^{\otimes N}$ as the marked entry. In a more general case, where the oracle function marks any other bit string, the $(N - 1)$-controlled-$Z$ gate should be executed in condition of the control qubits in the basis state matching the bit string. This can be implemented with an additional set of parallel $X$ gates before and after the regular $(N - 1)$-controlled-$Z$ gate.

For the qudit system, an oracle with a $(N - 1)$-controlled-$Z$ gate has a constant circuit depth of 1. In the conventional qubits basis, implementation of the oracle with an $(N - 1)$-controlled-$Z$ gate requires a circuit depth growing quadratically with the system size $N$ [13]. This can be reduced to a linear circuit depth growth of $8N - 20$ [33]. The addition of two sets of parallel $X$ gates to the oracle would add a constant depth of 2 for a qubit system, and a constant depth of 4 for a qudit system.

The output from the oracle module, is fed to Grover's Diffusion module with operational block diagram shown in figure 3. The diffusion operator increases the probability amplitude for the marked term by flipping the probability amplitudes of all terms in the superposition about the average. In a qudit system the diffusion operator grows linearly as $2N + 7$, with the two multi-action Hadamard gates as the driver in the linear term since the $X$ gates and $(N - 1)$-controlled-$Z$ gate are constant in depth (2 and 1 respectively). Operating in the qubit basis, the $H$ and $X$ gates have constant depths (2 and 1 respectively), but the module has a quadratic circuit depth due to the $(N - 1)$-controlled-$Z$ gate [13]. The diffusion module depth can be reduced to a linear depth of $8N - 14$, employing the same technique for the $(N - 1)$-controlled-$Z$ gate as in the oracle.

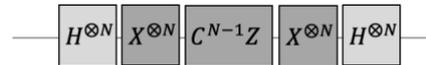

Figure 3. Generalized circuit implementation of Grover's Diffusion operator. For qudits of dimension $2^N$, Hadamard gates are applied in series and the set has a depth of N+1. The qudit NOT gates have a depth of 2 and the (N-1)-controlled-$Z$ gate has a depth of 1.

In order to increase the probability of measuring the marked term, Grover's Algorithm runs the superposition through the oracle and diffusion modules t times, where t $\propto O(\sqrt{D})$. As a result, the combined circuit depth for the oracle and diffusion modules contributes a multiple of t times to the total circuit depth required to run Grover's algorithm to maximum theoretical accuracy.

Finally, we compare the total circuit depth results for t iterations of Grover's algorithm for increasing qudit and qubit system size. The decomposition of the multiply-controlled-$Z$ gate into elementary gates results in a quadratic growth in depth for the qubit system. With the multiply-controlled-$Z$ gate appearing in both the oracle and the diffusion operator the total depth of the qubit circuit is dominated by this quadratic growth. Clearly, the circuit depth of a qubit system with quadratic growth will grow faster than a linear based qudit or qubit system. Instead, we will compare the two systems with linear circuit depth growth.

The total circuit depth for Grover's algorithm, with $(N-1)$-controlled-$Z$ gate as the oracle, using a qudit system grows linearly as $t(2N+8)+N+1$ for a given number of t iterations. For a qubit system, with a linear construction of the $(N-1)$-controlled-$Z$ gate, the overall circuit depth for t iterations grows as $t(16N-34)+2$. Table VI shows the total circuit depth for maximum theoretical accuracy of Grover's algorithm for a qudit system and a qubit system, where the gates are implemented by pulses ($U(2)$ rotations). As the system size increases, the circuit depth to run Grover's algorithm with high theoretical accuracy, qudit gates show a significant reduction in circuit depth compared to qubit gates.

| Equivalent N qubits (D = levels) | Iterations (t) | Theoretical Accuracy | Grover's Algorithm Total Depth (Pulses in Parallel) | |
|---|---|---|---|---|
| | | | Qudit | Qubits |
| 3 (D = 8) | 2 | 97.23% | 32 | 30 |
| 4 (D = 16) | 3 | 98.05% | 53 | 92 |
| 5 (D = 32) | 4 | 99.96% | 78 | 186 |
| 6 (D = 64) | 6 | 99.83% | 127 | 374 |
| 7 (D = 128) | 8 | 99.78% | 184 | 626 |

Table VI. Grover's Algorithm theoretical accuracy and total circuit depth for qubits and qudit gates implemented using pulses. Entries in white are the smallest number of iterations for first maximum in accuracy.

## IV. SUMMARY

We have presented a systematic and scalable construction of qudit-based quantum gates through a bijective mapping between the computation basis of N qubits and the $D = 2^N$ levels in a qudit of size $D+1$. The composition complexity of the qudit-based gates has been analyzed in terms of the total number of basic $U(2)$ rotations and the circuit depth. For the universal qubit gate set ($H$, $CNOT$, and $T$) implemented with a qudit the total number of rotations is linear to $D$ ($3D/4$, $3D/8$, and $3D/4$, respectively) while the circuit depth remains constant (2, 2, and 1, respectively) regardless of the system size. The number of the 2-dimentional subspaces specified by the qudit level pairing grows linearly with $D$ while the rotations in all orthogonal subspaces can be executed in parallel. In a qubit circuit, a single qubit gate can be applied to multiple qubits in parallel, while the complexity of the corresponding implementation in a qudit depends on the gate. The single qudit H gate equivalent to multiple qubit gate $H^{\otimes M}$, requires $D(2M+1)/4$ rotations with circuit depth $M+1$.

The complexity for multiple qubit $X$ gates, $X^{\otimes M}$, stays the same as a single qudit $X$ gate applying the same quantum logic. Qudit implementation greatly favors multiply-controlled gates, such as controlled-$NOT$ and controlled-$Z$. The depth remains constant (2 for controlled-$NOT$ and 1 for controlled-$Z$) independent of the system size. The required number of rotations decreases exponentially with the number of control qubits down to 2 for controlled-$NOT$ and 1 for controlled-$Z$ at $(N-1)$ control qubits.

Compared with qubit-based circuits, a qudit implementation is advantageous where the multiply controlled gates are employed. A significant circuit depth reduction has been shown in Grover's Algorithm. Using either the standard set of qubit elementary gates, or optimizing the circuit using linear depth multiply controlled-$Z$ gates, the depth of Grover's Algorithm grows quickly as the system size increases. Comparatively, the depth of Grover's Algorithm using qudits grows slowly as the system size increases, giving qudit circuits an advantage in depth over qubits. The scope of this paper does not include analysis of the effects of gate error. However, the reduction in the total number of gates leads to increase of the fidelity in general.

The development of physical quantum computing systems is focused on control of systems with larger numbers of computational qubits. The nature of quantum systems though is inherently more extensive than systems of two basis states, but this vast computational spaced is not utilized. As technology developments allow us greater control quantum systems, we may see development of multi-level qudit based quantum computers, which will allow for a greater computational efficiency in terms of number of rotation operations and circuit depth. In the meantime, qudit basis computation may find other applications where multiply controlled gates play an important role with the development of more advanced quantum algorithms. It will be of interest in future study to analyze the measurement schemes and consequence in single qudit, as well as gates and measurement in qubit-qudit or qudit-qudit systems.

[1] Arute, F., Arya, K., Babbush, R., Bacon, D., Bardin, J.C., Barends, R., Biswas, R., Boixo, S., Brandao, F.G., Buell, D.A. and Burkett, B., 2019. Quantum supremacy using a programmable superconducting processor. *Nature*, *574*(7779), pp.505-510.
[2] Gambetta, J., 2020. Ibm's roadmap for scaling quantum technology. *IBM Research Blog [Internet]*.
[3] Wright, K., Beck, K.M., Debnath, S., Amini, J.M., Nam, Y., Grzesiak, N., Chen, J.S., Pisenti, N.C., Chmielewski, M., Collins, C. and Hudek, K.M., 2019. Benchmarking an 11-qubit quantum computer. *Nature communications*, *10*(1), pp.1-6.
[4] Grzesiak, N., Blümel, R., Wright, K., Beck, K.M., Pisenti, N.C., Li, M., Chaplin, V., Amini, J.M., Debnath, S., Chen, J.S. and Nam, Y., 2020. Efficient arbitrary simultaneously entangling gates on a trapped-ion quantum computer. *Nature communications*, *11*(1), pp.1-6.
[5] Arrazola, J.M., Bergholm, V., Brádler, K., Bromley, T.R., Collins, M.J., Dhand, I., Fumagalli, A., Gerrits, T., Goussev, A., Helt, L.G. and Hundal, J., 2021. Quantum circuits with many photons on a programmable nanophotonic chip. *Nature*, *591*(7848), pp.54-60.
[6] Nielson, M.A. and Chuang, I.L., 2000. Quantum computation and quantum information.
[7] Kumar, P., 2013. Direct implementation of an N-qubit controlled-unitary gate in a single step. *Quantum information processing*, *12*(2), pp.1201-1223.
[8] Zhong, H.S., Wang, H., Deng, Y.H., Chen, M.C., Peng, L.C., Luo, Y.H., Qin, J., Wu, D., Ding, X., Hu, Y. and Hu, P., 2020. Quantum computational advantage using photons. *Science*, *370*(6523), pp.1460-1463.
[9] Ebadi, S., Wang, T.T., Levine, H., Keesling, A., Semeghini, G., Omran, A., Bluvstein, D., Samajdar, R., Pichler, H., Ho, W.W. and Choi, S., 2021. Quantum phases of matter on a 256-atom programmable quantum simulator. *Nature*, *595*(7866), pp.227-232.
[10] Dawson, C.M. and Nielsen, M.A., 2005. The solovay-kitaev algorithm. *arXiv preprint quant-ph/0505030*.
[11] Preskill, J., 2012. Quantum computing and the entanglement frontier. *arXiv preprint arXiv:1203.5813*.
[12] Preskill, J., 2018. Quantum computing in the NISQ era and beyond. *Quantum*, *2*, p.79.
[13] Barenco, A., Bennett, C.H., Cleve, R., DiVincenzo, D.P., Margolus, N., Shor, P., Sleator, T., Smolin, J.A. and Weinfurter, H., 1995. Elementary gates for quantum computation. *Physical review A*, *52*(5), p.3457.
[14] Maslov, D., Dueck, G.W. and Miller, D.M., 2007. Techniques for the synthesis of reversible Toffoli networks. *ACM Transactions on Design Automation of Electronic Systems (TODAES)*, *12*(4), pp.42-es.
[15] He, Y., Luo, M.X., Zhang, E., Wang, H.K. and Wang, X.F., 2017. Decompositions of n-qubit Toffoli gates with linear circuit complexity. *International Journal of Theoretical Physics*, *56*(7), pp.2350-2361.
[16] Shende, V.V. and Markov, I.L., 2008. On the CNOT-cost of TOFFOLI gates. *arXiv preprint arXiv:0803.2316*.
[17] Ferraro, E. and De Michielis, M., 2020. On the robustness of the hybrid qubit computational gates through simulated randomized benchmarking protocols. *Scientific Reports*, *10*(1), pp.1-10.
[18] Low, P.J., White, B.M., Cox, A.A., Day, M.L. and Senko, C., 2020. Practical trapped-ion protocols for universal qudit-based quantum computing. *Physical Review Research*, *2*(3), p.033128.
[19] Kiktenko, E.O., Nikolaeva, A.S., Xu, P., Shlyapnikov, G.V. and Fedorov, A.K., 2020. Scalable quantum computing with qudits on a graph. *Physical Review A*, *101*(2), p.022304.
[20] Dalla Chiara, M.L., Giuntini, R., Sergioli, G. and Leporini, R., 2018. A many-valued approach to quantum computational logics. *Fuzzy Sets and Systems*, *335*, pp.94-111.
[21] Kiktenko, E.O., Fedorov, A.K., Strakhov, A.A. and Man'Ko, V.I., 2015. Single qudit realization of the Deutsch algorithm using superconducting many-level quantum circuits. *Physics Letters A*, *379*(22-23), pp.1409-1413.
[22] Pavlidis, A. and Floratos, E., 2021. Quantum-Fourier-transform-based quantum arithmetic with qudits. *Physical Review A*, *103*(3), p.032417.
[23] Godfrin, C., Ferhat, A., Ballou, R., Klyatskaya, S., Ruben, M., Wernsdorfer, W. and Balestro, F., 2017. Operating quantum states in single magnetic molecules: implementation of Grover's quantum algorithm. *Physical review letters*, *119*(18), p.187702.
[24] Kues, M., Reimer, C., Roztocki, P., Cortés, L.R., Sciara, S., Wetzel, B., Zhang, Y., Cino, A., Chu, S.T., Little, B.E. and Moss, D.J., 2017. On-chip generation of high-dimensional entangled quantum states and their coherent control. *Nature*, *546*(7660), pp.622-626.
[25] Braumüller, J., Cramer, J., Schlör, S., Rotzinger, H., Radtke, L., Lukashenko, A., Yang, P., Skacel, S.T., Probst, S., Marthaler, M. and Guo, L., 2015. Multiphoton dressing of an anharmonic superconducting many-level quantum circuit. *Physical Review B*, *91*(5), p.054523.
[26] Lanyon, B.P., Barbieri, M., Almeida, M.P., Jennewein, T., Ralph, T.C., Resch, K.J., Pryde, G.J., O'brien, J.L., Gilchrist, A. and White, A.G., 2009. Simplifying quantum logic using higher-dimensional Hilbert spaces. *Nature Physics*, *5*(2), pp.134-140.
[27] Luo, Y.H., Zhong, H.S., Erhard, M., Wang, X.L., Peng, L.C., Krenn, M., Jiang, X., Li, L., Liu, N.L., Lu, C.Y. and Zeilinger, A., 2019. Quantum teleportation in high dimensions. *Physical review letters*, *123*(7), p.070505.
[28] Bocharov, A., Roetteler, M. and Svore, K.M., 2017. Factoring with qutrits: Shor's algorithm on ternary and metaplectic quantum architectures. *Physical Review A*, *96*(1), p.012306.
[29] Kiktenko, E.O., Fedorov, A.K., Man'Ko, O.V. and Man'Ko, V.I., 2015. Multilevel superconducting circuits as two-qubit systems: Operations, state preparation, and entropic inequalities. *Physical Review A*, *91*(4), p.042312.
[30] Ivanov, S.S., Tonchev, H.S. and Vitanov, N.V., 2012. Time-efficient implementation of quantum search with qudits. *Physical Review A*, *85*(6), p.062321.

[31] O'Leary, D.P., Brennen, G.K. and Bullock, S.S., 2006. Parallelism for quantum computation with qudits. *Physical Review A*, *74*(3), p.032334.
[32] Muthukrishnan, A. and Stroud Jr, C.R., 2000. Multivalued logic gates for quantum computation. *Physical review A*, *62*(5), p.052309.
[33] Saeedi, M. and Pedram, M., 2013. Linear-depth quantum circuits for n-qubit Toffoli gates with no ancilla. *Physical Review A*, *87*(6), p.062318.
[34] Nielson, M.A. and Chuang, I.L., 2000. Quantum computation and quantum information, p.251.
[35] Gidney, C., 2015. Using quantum gates instead of ancilla bits. Available at: http://algassert.com/circuits/2015/06/22/Using-Quantum-Gates-instead-of-Ancilla-Bits.html (Accessed: 16-07-2021).
[36] Zhang, K. and Korepin, V.E., 2020. Depth optimization of quantum search algorithms beyond Grover's algorithm. *Physical Review A*, *101*(3), p.032346.
[37] Wang, Y. and Krstic, P.S., 2020. Prospect of using Grover's search in the noisy-intermediate-scale quantum-computer era. *Physical Review A*, *102*(4), p.042609.
[38] Mandviwalla, A., Ohshiro, K. and Ji, B., 2018, December. Implementing Grover's algorithm on the IBM quantum computers. In *2018 IEEE International Conference on Big Data (Big Data)* (pp. 2531-2537). IEEE.